# An Analysis of the Long-term Photometric Behavior of ε Aurigae

**Brian K. Kloppenborg**
*Department of Physics and Astronomy, University of Denver, 2112 East Wesley Avenue, Denver, CO 80208; bkloppen@du.edu*

**Jeffrey L. Hopkins**
*Hopkins Phoenix Observatory, 7812 West Clayton Drive, Phoenix, AZ 85033; phxjeff@hposoft.com*

**Robert E. Stencel**
*University of Denver, Department of Physics and Astronomy, 2112 E. Wesley Avenue, Denver, CO 80208*; rstencel@du.edu

## Abstract

The lure of a 50% reduction in light has brought a multitude of observers and researchers to ε Aur every twenty-seven years, but few have paid attention to the system outside of eclipse. As early as the late 1800s, it was clear that the system undergoes some form of quasi-periodic variation outside of totality, but few considered this effect in their research until the mid-1950s. In this work we focus exclusively on the out-of-eclipse (OOE) variations seen in this system. We have digitized twenty-seven sources of historic photometry from eighty-one different observers. Two of these sources provide twenty-seven years of inter-eclipse *UBV* photometry which we have analyzed using modern period finding techniques. We have discovered the F-star variations are multi-periodic with at least two periods that evolve in time at $\Delta P \approx -1.5$ day/year. These periods are detected when they manifest as near-sinusoidal variations at 3,200-day intervals. We discuss our work in an evolutionary context by comparing the behavior found in ε Aur with bona-fide supergiant and post-AGB stars of similar spectral type. Based upon our qualitative comparison, we find the photometric behavior of the F-star in the ε Aur system is more indicative of supergiant behavior. Therefore the star is more likely to be a "traditional supergiant" than a post-AGB object. We encourage continued photometric monitoring of this system to test our predictions.

## 1. Introduction

ε Aurigae is a 27.1-year single line eclipsing spectroscopic binary that has long been wrapped in an enigma. Ever since a dimming of the system was discovered by Fritsch (1824), and its periodicity established (Ludendorff 1903), astronomers have speculated about the cause of this variation. Based upon radial velocity measurements, ε Aur was classified as a spectroscopic binary (Vogel 1903), which hinted that the dimming might be an eclipse. Later application of Henry Norris Russel's binary star theory (Russell 1912a, 1912b) came to a striking revelation: the companion to the F-type supergiant was nearly equal in mass, yet spectroscopically invisible (Shapley1915).

Over the next decade several theories were advanced to explain this startling conclusion (for example, Ludendorff 1912; Kuiper *et al.* 1937; Schoenberg and Jung 1938; Kopal 1954); however, it was Huang (1965) who proposed that the eclipse was caused by a disk of opaque material that enshrouded the secondary component. Although Huang's analytic model replicated the eclipse light curve, the disk theory remained unproven until Backman (1985) detected an infrared excess that corresponded to a 500 K blackbody source. Recently, the disk theory was vindicated by interferometric observations of the eclipse. These images show the F-star is partially obscured by a disk of opaque material (Kloppenborg *et al.* 2010, 2011) which is responsible for the dimming observed photometrically.

Although most research on this system concentrated on the eclipse itself, a few works have looked at the system outside of eclipse. Since the early 1900s, it has been known that the F-star exhibits 0.1-magnitude variations in *V*-band outside of eclipse. Indeed, the earliest discussion of these out-of-eclipse (OOE) variations were by Shapley (1915, p. 20). He comments on variations in visual photometry with an amplitude of $\Delta$Vis = 0.3 magnitude. Because observational errors were not fully characterized, Shapley treated these results with caution. Later, Güssow (1928) spotted a $\Delta$Vis = 0.15-magnitude variation outside of eclipse that was corroborated by two photoelectric photometers (Shapley 1928), thereby confirming the presence of the OOE variations. Shapley (1928) concluded that these variations arose from a ~355-day quasi-periodic variation; however, the exact period was poorly constrained by these data.

After the 1983–1985 eclipse, Kemp *et al.* (1986) proposed that a ~100-day period may exist in polarimetry data. Later, Henson (1989) showed there was little to no wavelength dependence in the variations, implying that the source of polarization is Thompson scattering from free electrons. In his dissertation, Henson found intervals where there were variations in Stokes Q, but little to nothing in Stokes U. This was interpreted to be caused by the F-star having two major axes for polarization, inclined at an angle of 45 degrees with respect to each other. Like many of the other studies, a visual inspection of Henson's data showed that some long trends may indeed exist, but nothing was strictly periodic. From the post-eclipse polarimetry, Henson concluded that the photometric and polarimetric variations might be caused by non-radial pulsation in low-order $\ell = 1,2$ $m = \pm 1$ modes. This notion is supported by the recent automated classification of ε Aur as an α Cyg variable (Dubath *et al.* 2011a).

After the 1985 eclipse, many authors sought to determine periods of the OOE variation. Using data from the first five years after the eclipse, Nha *et al.* (1993) found occasional stable variational patterns would set in (in particular around JD 2447085–2447163) with $\Delta U = 0.27$, $\Delta B = 0.17$, and $\Delta V = 0.08$, and a characteristic period of 95.5 days. Later, Hopkins and Stencel (2008) analyzed their inter-eclipse V-band photometric data using the PERANSO software package (Husar 2006). They found two dominant peaks in the Fourier power spectra with 65- and 90-day periods.

Perhaps the most comprehensive period analysis effort was made by Kim (2008). They used the CLEANest Fourier transform algorithm (Foster 1995) and the Weighted Wavelet Z-transform (WWZ; Foster 1996) on nearly 160 years of photometry of ε Aur. Using these two algorithms they identified several periods which led them to conclude ε Aur may be a double or multi-periodic pulsator.

Here we extend the work of our predecessors by analyzing twenty-seven-years worth of inter-eclipse *UBV* photometry. In the following sections we discuss our sources of data, our analysis methods, and the results. We then conclude with a discussion of our results in a stellar evolution context.

## 2. Data sources

### 2.1. Historic photometry

ε Aur has a rich history of photometric observations. We conducted a comprehensive literature review and found twenty-seven sources of photometry from eighty-one different observers. We digitized all of these data and intend to submit them to the AAVSO International Database or VizieR (when copyright allows) after the publication of this article. A full discussion of the data sources, assumed uncertainties, and digitization methods is in Kloppenborg (2012); however, we have summarized the important aspects of these data in Table 1.

### 2.2. Phoenix-10

The Phoenix-10 Automated Photoelectric telescope, designed by Louis Boyd, obtained a total of 1,570 *U*-, 1,581 *B*-, and 1,595 *V*-band observations of ε Aur between 1983 and 2005. The system consisted of a 1P21 photomultiplier mounted on a 10-inch $f/4$ Newtonian telescope. A detailed description of this setup can be found in Boyd *et al.* (1984b). The telescope was originally located in downtown Phoenix, Arizona, until it was moved to Mount Hopkins during the summer of 1986. The system was then moved to Washington Camp in Patagonia, Arizona, in 1996 where it operated until 2005.

The earliest data on ε Aur were obtained in 1983 November and covered most of the 1983–1984 eclipse (Boyd *et al.* 1984a). These data cover intervals JD 2445646–2455699 (Breger 1982, file 131), JD 2445701–2445785 (Breger 1985, file 136), and JD 2445792–2445972 (Breger 1988, file 137) (1983 November 3–1984 September 29). Although the photometer collected data

between 1984 September and 1987 September, the original data were lost due to a hardware failure (Boyd 2010). As far as we have been able to determine, these data were not published in subsequent issues of the *IAU Archives of Unpublished Observations of Variable Stars*. In addition to these data, our publication includes unpublished data from the APT-10 which starts on JD 2447066 (1987 September 27) and ends on 2453457 (2005 March 27).

The standard observing sequence for ε Aur was KSCVCVCVCSK (K=Check, S=Sky, C=Comparison, V=Variable) iterating through the *UBV* filters (see Boyd *et al.* 1984b, Table 1) with 10-second integrations. Information on the target, check, and comparison stars are summarized in Table 2. These differential measurements were corrected for extinction and transformed into the standard *UBV* system. The automated reduction pipeline discarded any observations with an internal standard error of ± 20 milli-magnitudes or greater. Typical external errors are ± 0.011, ± 0.014, ± 0.023 magnitude in *V*, *B*, and *U* filters, respectively, with mean internal errors of ± 0.005, ± 0.005, and ± 0.009 (Strassmeier and Hall 1988). The stability of the system has been satisfactory on decade-long timescales (Hall and Henry 1992; Hall *et al.* 1986).

2.3. Hopkins *UBV*

Co-author Jeffrey Hopkins collected 811 *U*, 815 *B*, and 993 *V* differential magnitudes at the Hopkins Phoenix Observatory (HPO) in Phoenix, Arizona. The data consist of two large blocks: the first began on 1982 September 09 and ended on 1988 December 23, and the second series started on 2003 December 04 and ended on 2011 April 25. At this time the photometric program at HPO ended.

The HPO setup consisted of a 1P21 photomultiplier mounted to an 8-inch Celestron C-8 telescope with standard Johnson *UBV* filters. Observations of ε Aur were conducted in CSVSCSVSCSVSCS format, each composed of three 10-second integrations in one of the three filters. Nightly extinction coefficients were determined and applied. Color correction was determined on a monthly basis. After the 1980 observing season, λ Aurigae was used as the sole comparison star. The assumed magnitudes for λ Aurigae were *V* = 4.71, (*B–V*) = 0.63, and (*U–B*) = 0.12. A subset of these data have been discussed in Chadima *et al.* (2011).

2.4. AAVSO Bright Star Monitor

Starting on 2009 October 16, ε Aur was placed on the American Association of Variable Star Observers' Bright Star Monitor (BSM) observing program. The BSM consisted of a Takahashi FS-60CB with a field flattener; 60-mm *f*/6.2 telescope and a SBIG ST-8XME camera with Johnson/Cousins $BVR_cI_c$ and clear filters. Data were reduced by Arne Henden at AAVSO headquarters to the standard photometric system. All color and extinction corrections were applied before data were submitted to the AAVSO International Database. Ongoing observations from this instrument can be found in the AAVSO database under the observer code "HQA."

2.5. Solar mass ejection imager

For the sake of completeness, we also mention our work on data from the Solar Mass Ejection Imager (SMEI; Simnett *et al.* 2003). Although SMEI's primary mission is to map the large-scale variations in heliospheric electron densities by observing Thompson-scattered sunlight, it also collected precision photometry on ~20,000 stars with $V < 8$. Through each 102-minute orbit, most regions in the sky were covered by a dozen or more frames through one of SMEI's three baffled, unfiltered CCD cameras.

The instrument was designed for 0.1% photometry and, when proper photometric extraction is performed, this precision was realized on stars brighter than fifth magnitude. On average the uncertainty on fainter stars was proportionally worse by the ratio of the star's brightness to fourth magnitude, meaning faint eighth magnitude stars still have better-than-ground photometry precision. The instrument was operational from launch until September 2011 when it was deactivated due to budget constraints. Our work with these data will be discussed in a future publication.

# 3. Analysis

The sources of photometry listed above were very inhomogeneous; consisting of multiple filters, reduction methods, observatories, instruments, and even reference star magnitudes. We created a script that finds overlaps between two data sets of the same filter, bins the data, and then calculates the coefficients required to scale/offset the data to the same quasi-system using a weighted least-squares technique using the following equation:

$$Ai = a + b B_j + c t_j \qquad (1)$$

Here $A_i$ is the $i^{th}$ entry in the reference photometry set, $B_j$ is the $j^{th}$ entry in the comparison photometry data set occurring at time $t_j$, $a$ is a zero-point offset, $b$ accounts for non-Pogson magnitudes (present in only our earliest visual photometric data), and $c$ corrects for a time-dependent drift of the comparison photometry set. In all of the filtered photometry, only $a$ was required. In the visual data, $a$ and $b$ were required. As our work here is concerned with the variations, rather than absolutely calibrated photometry, we selected Hopkins *U*, BSM *B*, and BSM *V* as references. Offsets (*a*) between various photometric sources are summarized in Table 3.

After the offsets were determined we passed the data through the Windows implementation of WWZ, WINWWZ. Here we have used data from the range JD 2446000–2455000 in 10-day steps with $f_{low} = 0.004$, $f_{high} = 0.05$, and $\Delta f = 0.0001$. Our WWZ decay constant was 0.0125.

## 4. Results

In Figure 1 we plot the twenty-seven-years worth of inter-eclipse *UBV* photometry from the APT-10 and HPO observatories. The V data have been plotted unaltered, but *U* have been offset by –1.3 magnitude and *B* by –0.8 magnitude. Internal photometric uncertainties (~ 10 milli-magnitudes) are about the thickness of the lines. Each observing season consists of approximately 200 days of data with bi-nightly sampling followed by 165-day gaps where the star was not visible at night. All of the data were corrected for extinction and transformed to the standard *UBV* photometric system. A visual inspection of the data reveals no single consistent period is present.

Historically, there have been several reported instances of short-term (that is, few-day long) events which we suspect are flares. Albo and Sorgsepp (1974) reported a $\Delta U = 0.2$, $\Delta B = 0.1$, and $\Delta V = 0.06$ brightening that lasted five days around JD 2439968 (1968 April 21). Similarly, Nha and Lee (1983) noted a rapid (few hour) 0.4 magnitude rise in the blue filter and 0.2 magnitude in the yellow filter on JD 2445356 (1983 January 21). We have noticed a two-night flare in our data set starting on JD 2446736 (1986 November 01). This event resulted in a $\Delta U = 0.2$, $\Delta B = 0.1$, and $\Delta V = 0.7$ photometric increase, strangely opposite of historic records. Continued *UBV* photometry appears to be a efficient way of detecting these events.

In Figure 2 we plot the current eclipse light curve in *UBVRIJH* filters from the above sources and the AAVSO database. We note that the eclipse appears to be slightly wavelength-dependent, particularly from mid-eclipse to third contact. This can be seen from the downward slope of the *U*-band data and flat trend in *H*-band. We suggest this is due to additional small particles coming into the line of sight from a sublimation zone on the F-star facing side of the disk.

In Figures 3 through 5 we plot the WWZ results for the *U*, *B*, and *V* filters, respectively. The color in these figures is the WWZ output with higher value indicating stronger presence of that particular period. Because the APT-10 was not operational during the interval JD 2449500–JD 245000, we blocked out the WWZ result in this region.

From inspection of these figures, it is clear that no single period can accurately describe the variations seen in ε Aur. In Table 4 we list peaks whose WWZ coefficient was greater than 100. In the *U*-band data the two most dominant periods are centered at (8977, 102.7) and (12259, 87.9) (henceforth the "upper track"). These peaks were spaced apart by 3,282 days with $\Delta P = -14.8$ days. When the WWZ output was high, the radial velocity (see Stefanik *et al.* 2010 for data) and photometric changes were in phase. The combination of these two effects lead us to believe these events should be regarded as significant.

Operating on the assumption that these peaks provided a glimpse of period evolution in the F-star, we intentionally sought variations following a parallel evolutionary path. Three peaks located at (7166, 90), (10492, 82.7), and (13744, 68.9) (hereafter the "lower track") followed a similar evolution. Like the "upper track," these peaks were separated by nearly 3,300 days.

Peaks at these locations also appeared in the *B*- and *V*-band data, although with much lower significance. The WWZ value in WINWWZ was determined by a $\chi^2$-like metric that does not consider the uncertainties in the data. Therefore the additional scatter seen in adjacent *B* or *V* photometry, although within uncertainties, results in a lower WWZ value. The higher amplitudes and greater night-to-night consistency causes the *U* data to have larger WWZ values, on average, than the *B* and *V* data.

In the *B*-band WWZ, we see a few peaks in the 80- to 100-day range appearing from time to time, although they are clearly not stable. Likewise, in the *V*-band WWZ there is a single dominant peak of (6100, 90), but otherwise little hint of a stable variational pattern. We do not suggest the WWZ <100 results should be given much consideration; however, in the range JD 24450000–24455000 there were several commensurate periods that appeared to evolve downward at a rate of several days/year.

In Table 5 we used the above observations as a guide and predicted dates when stable pulsational patterns should develop and their periods. We note that the 118-day period around JD 2445695 (near third contact of the 1984 eclipse) did not manifest; however, near the end of the 2009–2011 eclipse a sawtooth-like pattern with a 61- to 76-day period developed (see Figure 2). This is tantalizingly close to the ~73-day period we predicted would develop at this time.

To test our extrapolation, we attempted a WWZ analysis on the visual data. The only set that spanned an entire inter-eclipse interval was collected by Plassman (Güssow1936). Data within the interval 2422000–2428000 (see Figure 6) clearly show the presence of the OOE variations with characteristic periods of 330–370 days. This value was nearly 100 days longer than what we predict for this time interval, implying our extrapolation should not be regarded as highly predictive until further period characterization is conducted.

## 5. Discussion and conclusion

We have created the first long-term *UBV* photometric record of ε Aur using the data from the APT-10 and HPO observatories. These data show stable variational patterns developed on 3,200-day timescales. In the *U*-band WWZ output, we have identified two parallel tracks of stable variations that evolve at a rate of $\Delta P = -1.6$ day/year and $\Delta P = -1.2$ day/year for the "upper" and "lower" track, respectively. Extrapolating these results, we have identified dates at which we anticipate stable variational patterns will manifest and predicted the periods that they should have. Our extrapolation to JD 2455541 predicts a 73-day period should have developed; it is tantalizingly close to the 61- to 76-day period that was observed during the second-half of the 2009–2011 eclipse. Our interpretation below is based on this "two-track" notion and likely underestimates the true complexity in this system. In Table 4 we provide all peaks with a WWZ > 100 for the *UBV* photometry in hopes that they will be useful to future researchers.

At the time of this writing, a consistent asteroseismic interpretation for evolved supergiant-class stars does not exist due to the uncertainties underlying the theoretical calculations of mixing theory and radiation pressure (Aerts *et al.* 2010, ch. 2). Therefore, we cannot provide a rigorous, quantitative interpretation of the periods which we have observed. Instead, we interpret the observed periods qualitatively by comparing them with observed supergiant and post-AGB behavior. Regrettably, few comparative studies of F-type supergiant/post-AGB stars exist, especially multi-decade surveys. Therefore our interpretations are inherently biased. We have attempted to discuss these biases throughly and indicate where our study could benefit from future research.

It had been known for some time that stars near the F0Ia spectral and luminosity class show low-level variations with 0.015–0.025 magnitude amplitudes in the V-band (Maeder 1980). An investigation by van Leeuwen *et al.* (1998) of twenty-four super- and hyper-giant stars from the LMC, SMC, and the Milky Way using HIPPARCOS photometry showed that all of these B to late-G stars exhibited photometric changes that were not strictly periodic. Indeed, many of these "periods" would be better described as "quasi-" or "pseudo-periods." Across this region of the HR diagram, stars tended to show variations on 10- to 100-day timescales.

Indeed, in this respect ε Aur could easily be regarded as a recently-evolved supergiant. Several stars in the van Leeuwen *et al.* (1998) sample were a close match for ε Aur: At a slightly higher temperature, HD 269541 (HIP 25448, A8:Ia+, in the LMC) has $\Delta V_T = 0.1$-magnitude variations with several short periods in the 8- to 40-day interval, and two longer periods at 146 and 182 days. The slightly cooler HD 269697 (HIP 25892, F5Ia, in the LMC) had two equally significant periods at 48 and 84 days with photometric variations between 0.01 and 0.05 magnitude in $\Delta V_T$. HD 74180 (HIP 42570, F2Ia, in the Milky Way) was the closest match to ε Aur in the van Leeuwen *et al.* (1998) sample. This star showed quasi-periods at 53, 80, and 160 days with variations of 0.06 magnitude. It is also worth mentioning that an automated photometric classification scheme considered ε Aur to be an α Cyg variable (Dubath *et al.* 2011), a class of luminous supergiants undergoing non-radial pulsation.

Post-AGB stars make up a very heterogeneous class of objects, therefore it is difficult to discuss their properties, let alone discuss any reasons for membership (or lack thereof) for one particular star. We have searched the "Torun catalogue of Galactic post-AGB and related objects" (Szczerba *et al.* 2007) for systems of similar spectral type and identical luminosities to ε Aur. The best-match is AR Pup (F0Iab, HIP 39376) which features multi-periodic (RVb) pulsation with $\Delta V = 0.5$ and timescales of $76.4 \pm 4$- and $1{,}250 \pm 300$- day periods (Kiss *et al.* 2007). Although the timescales match, the variational pattern (that is, highly predictable, stable) does not match what is seen in ε Aur. Likewise the pattern seen in the cooler V340 Ser (HD 158616, F8) has similar timescales (87.7d and 131d, Arkhipova *et al.* 2011), but is obviously multi-periodic and easily predicted. To complete our view of variations seen around the F0Ia class, long-term photometric studies of the post-AGB stars IRAS 10197-5750 (A2Iab:e, 2MASS J10213385-5805476), IRAS 16206-5956 (A3Iab:e, 2MASS J16250261-6003323), IRAS 06530-0213 (F0Iab, 2MASS J06553181-0217283), HD 101584 (F0Iabpe, HIP 56992), and HD 187885 (F2/F3Iab, IRAS 19500-1709) would be beneficial.

Aggregate statistics of post-AGB stars imply systems of similar spectral types to ε Aur have significantly shorter periods than their supergiant counterparts. For example, Hrivnak *et al.* (2010) studied a series of C-rich post-AGB stars and found a strong correlation between the effective temperature and period. The relationship predicts that higher-temperature post-AGBs will have shorter periods following a linear trend: $\Delta P / \Delta T_{eff} = -0.047$ day K[1]. Their Figure 18 suggested ε Aur should exhibit variations with a ~40-day timescale, a factor of 1.7 less than what we have observed. Their work on O-rich stars appears to be forthcoming (see Shaw *et al.* 2011). In a sample of five post-AGB stars Arkhipova *et al.* (2011) found a similar trend. Their Figure 8 predicts periods of ~65 days, a factor of 1.25 to 1.5 shorter than what we have observed. A majority of their program stars also were multiperiodic, with ratios of $P_1/P_2$ ~1.03 to 1.09, whereas ε Aur shows a higher ratio of 1.24 to 1.27.

Until this point we have compared the variational patterns in ε Aur against single stars. As noted above, the stable variation patterns develop at 3,200-day intervals, which is nearly 1/3 of the 27.1-year (9,890-day) orbital period. It would appear the companion is influencing the pulsational properties of the F-star. As the orbit is eccentric ($e$ ~0.227 or $e$ ~0.249–0.256, Stefanik *et al.* 2010, Chadima *et al.* 2010, respectively) one might anticipate tidal flows to be induced in the F-star's tenuously-bound atmosphere (log $g$ ~1, Sadakane *et al.* 2010) during periastron passage; however, dissipation timescales would certainly be less than the nine-year interval seen between stable variational patterns (see Moreno *et al.* 2011, and references therein for a discussion of the theoretical framework). Instead we speculate that gravitational forcing due to orbital motion is exciting natural resonant frequencies in the F-star (this theory is shown to be possible in main sequence objects—see Goldreich and Nicholson 1989; Rocca 1989; Witte and Savonije 1999a, 1999b; Zahn 1975, 1977). This conjecture predicts that the excitations should repeat at the same orbital phases and is therefore testable by continued photometric monitoring. The next dates when such events might happen are JD ~2457000 (2014 December) and JD ~2457000 (2019 December). The development of a consistent asteroseismic theory for supergiants may provide an earlier test of our hypothesis.

Given this information and the photometric behavior discussed above,

we consider it unlikely that the F-star is a post-AGB object and conclude, on a qualitative basis, that the F-star is more likely a traditional supergiant. Implications for the evolutionary state and physical properties of the disk will be discussed in forthcoming publications.

## 6. Acknowledgements

Participants from the University of Denver are grateful for the bequest of William Hershel Womble in support of astronomy at the University of Denver. They acknowledge support from National Science Foundation through ISE grant DRL-0840188 (Citizen Sky) to the American Association of Variable Star Observers and AST grant 10-16678 to the University of Denver. We acknowledge with thanks the variable star observations from the AAVSO International Database contributed by observers worldwide and used in this research. This research has made use of NASA's Astrophysics Data System Bibliographic Services and the SIMBAD database, operated at CDS, Strasbourg, France.

Table 1. All sources of photometry (historic and new).

| Source | Observer[a] | JD Start | JD End | Obs. | Type[b] | Notes |
|--------|-----------|----------|--------|------|---------|-------|
| Ludendorff (1903) | Argelander | 2393950 | 2404492 | 56 | St | |
| Ludendorff (1903) | Heis | 2394511 | 2400252 | 78 | St | |
| Ludendorff (1903) | Oudemains | 2398490 | 2399058 | 16 | St | |
| Ludendorff (1903) | Argelander | 2403950 | 2404492 | 56 | St | |
| Ludendorff (1903) | Schoenfeld | 2404049 | 2405988 | 29 | St | |
| Ludendorff (1903) | Schwab | 2406614 | 2416245 | 65 | St | |
| Ludendorff (1903) | Plassman | 2408093 | 2414961 | 85 | St | |
| Ludendorff (1903) | Sawyer | 2409147 | 2413645 | 50 | St | |
| Ludendorff (1903) | Porro | 2411360 | 2411391 | 3 | St | |
| Ludendorff (1903) | Luitzet | 2414288 | 2416217 | 55 | St | |
| Ludendorff (1903) | von Prittwitz | 2414584 | 2416080 | 18 | Pm | |
| Ludendorff (1903) | Plassman | 2415259 | 2416223 | 29 | St | |
| Ludendorff (1903) | Kopff | 2415709 | 2415825 | 6 | St | |
| Ludendorff (1903) | Goetz | 2416107 | 2416196 | 6 | St | |
| Ludendorff (1912) | | 2395304 | 2406567 | 109 | St | |
| Wendell (1913) | | 2416574 | 2419706 | 41 | St | |
| Shapley (1928) | | 2416574 | 2419691 | 26 | Pe | |
| Huffer (1932) | | 2425267 | 2426467 | 98 | Pe | |
| Ivanov and Scharbe (1934) | | 2417821 | 2426104 | 84 | St | |
| Danjon (1936) | | 2425130 | 2429238 | 97 | St | |
| Güssow (1936) | Nijland (1) | 2413488 | 2414280 | 22 | St | |
| Güssow (1936) | Plassman | 2416380 | 2427785 | 430 | St | |

*Table continued on next page*

Table 1. All sources of photometry (historic and new), cont.

| Source | Observer[a] | JD Start | JD End | Obs. | Type[b] | Notes |
|--------|-----------|----------|--------|------|--------|-------|
| Güssow (1936) | Enebo | 2416425 | 2417326 | 31 | St | |
| Güssow (1936) | Wendell | 2416574 | 2419691 | 25 | Pe | |
| Güssow (1936) | Schiller | 2416848 | 2416973 | 12 | St | |
| Güssow (1936) | Lohnert | 2417326 | 2417498 | 5 | St | |
| Güssow (1936) | Scharbe (1) | 2417824 | 2418653 | 12 | St | |
| Güssow (1936) | Horning | 2418192 | 2420589 | 51 | St | |
| Güssow (1936) | Mundler | 2418323 | 2418657 | 13 | St | |
| Güssow (1936) | Lau | 2419275 | 2429779 | 16 | St | |
| Güssow (1936) | Menze | 2420031 | 2420958 | 39 | St | |
| Güssow (1936) | Guthnick | 2420147 | 2420175 | 10 | Pe | |
| Güssow (1936) | Johasson | 2422687 | 2423374 | 12 | St | |
| Güssow (1936) | Guthnick and Pavel | 2422940 | 2423361 | 27 | Pe | |
| Güssow (1936) | Gadomski | 2423016 | 2427060 | 31 | St | |
| Güssow (1936) | Graff | 2424251 | 2425957 | 13 | Pm | |
| Güssow (1936) | Kordylewski | 2424647 | 2425953 | 29 | St | |
| Güssow (1936) | Güssow | 2424808 | 2427762 | 145 | Pm | |
| Güssow (1936) | Kukarkin | 2425100 | 2426242 | 34 | St | |
| Güssow (1936) | Beyer | 2425126 | 2426196 | 44 | St | |
| Güssow (1936) | Danjon | 2425139 | 2426636 | 34 | Pm | |
| Güssow (1936) | Jacchia | 2425177 | 2426210 | 41 | St | |
| Güssow (1936) | Pagaczewski | 2425185 | 2426034 | 13 | St | |
| Güssow (1936) | Scharbe (2) | 2425237 | 2426104 | 15 | St | |

Table 1. All sources of photometry (historic and new), cont.

| Source | Observer[a] | JD Start | JD End | Obs. | Type[b] | Notes |
|--------|-------------|----------|--------|------|---------|-------|
| Güssow (1936) | Stebbins and Huffer | 2425267 | 2426467 | 98 | Pe | |
| Güssow (1936) | Tschernov | 2425296 | 2426090 | 25 | St | |
| Güssow (1936) | Mrazek | 2425322 | 2425857 | 4 | Pm | |
| Güssow (1936) | Nijland (2) | 2425322 | 2426436 | 40 | St | |
| Güssow (1936) | Dziewulski | 2425364 | 2426096 | 22 | St | |
| Güssow (1936) | Menze2 | 2425529 | 2426065 | 23 | St | |
| Güssow (1936) | Kopal | 2425925 | 2426465 | 18 | St | |
| Emberson et al. (1938) | | 2428903 | 2428920 | 7 | Broadband-IR | Not used |
| Thiessen (1957) | | 2435374 | 2435942 | 10 | yellow,blue,violet | |
| Larsson-Leander (1959) | | 2435428 | 2436334 | 121 | PV | |
| Widorn (1959) | | 2435128 | 2436295 | 122 | yellow,blue | Not used |
| Fredrick (1960) | | 2434761 | 2435903 | 143 | Pe, 4 color | Not transformed |
| Larsson-Leander (1962) | | 2436457 | 2437023 | 52 | PV | |
| Kopylov and Kumaigorodskaya (1963) | | 2435111 | 2436695 | 251 | PV | |
| Mitchell (1964) | | | | | UBVRIJHKLN | Not used |
| Low and Mitchell (1965) | | | | | UBVRIJHKLN | Not used |
| Stub (1972) | | 2438755 | 2439925 | 220 | Pe, 10 color | |
| Albo and Sorgsepp (1974) | | 2439947 | 2439988 | 144 | UBV | |
| Albo (1977) | | 2440460 | 2440548 | 42 | UBV | |
| Breger (1982, 1985, 1988) | | 2445646 | 2445972 | 453 | UBV | Boyd, IAU |
| JAPOA (1983) | | 2445239 | 2445409 | 123 | UBV | |

*Table continued on next page*

Table 1. All sources of photometry (historic and new), cont.

| Source | Observer[a] | JD Start | JD End | Obs. | Type[b] | Notes |
|---|---|---|---|---|---|---|
| Backman et al. (1984) | | 2444269 | 2445382 | 107 | JHKL'MNQ | |
| Bhatt et al. (1984) | | 2445044 | 2445798 | 84 | BVRIJHK | |
| Hopkins UBV (1) | | 2445222 | 2447520 | 1140 | UBV | |
| Flin et al. (1985) | | 2445065 | 2445937 | 330 | UBV | |
| Parthasarathy and Frueh (1986) | | 2445208 | 2445426 | 983 | UBV,uvby | |
| Boyd UBV | | 2447066 | 2453457 | 4746 | UBV | Unpublished |
| Chochol and Žižovský (1987) | | 2445043 | 2445664 | 207 | UBV | |
| Strassmeier et al. (1999) | | 2450396 | 2450427 | 216 | by | |
| Taranova and Shenavrin (2001) | | 2445032 | 2451652 | 243 | UBVRJHKLM | Download from CDS |
| Hopkins UBV (2) | | 2452979 | 2455678 | 1836 | UBV | |
| AAVSO BSM | | 2455122 | present | 960 | BVRI | |

[a] *Identfied only if from a compilation source.* [b] *The photometric system used or one of the following:* St = *Step magnitudes (mechanically assisted visual data), Pm* = *Photometric, Pe = photoelectric, and PV = photovisual. See original reference for further details.*

Table 2. Star information for Boyd photometric data.

| Object | R. A. | | | Dec. | | | Epoch | Role* | Other Names |
|--------|---|---|---|---|---|---|-------|-------|-------------|
| | h | m | s | ° | ' | " | | | |
| HD 34411 | 05 | 19 | 08 | 40 | 05 | 57 | 2000 | K | HR 1729, SAO 40233 |
| HD 32655 | 05 | 06 | 50 | 43 | 10 | 29 | 2001 | C | HR 1644, SAO 40029 |
| HD 31964 | 05 | 01 | 58 | 43 | 49 | 24 | 2002 | V | HR 1605, SAO 39955 |
| Sky | 05 | 04 | 24 | 43 | 29 | 57 | 2003 | S | |

*Role: K = Check, S = Sky, C = Comparison, V = Variable*

Table 3. Offsets between observers. "ref" indicates this was the reference photometry set for the associated column.

| Source | U | B | V |
|--------|---|---|---|
| 1983 Boyd | 6.944964 | 6.704575 | 5.955698 |
| 1984 Hopkins | −0.022894 | −0.056568 | −0.075453 |
| 1987 Boyd | −0.124972 | −0.007964 | −0.043503 |
| 2011 Hopkins | ref | −0.045311 | −0.044434 |
| 2011 AAVSO BSM | N/A | ref | ref |

Table 4. Peak periods observed in the UBV WWZ transforms roughly grouped by date. Dates have been rounded to the nearest 10. Periods and WWZ output are rounded to integer values. MJD = JD − 2440000.

| | U | | | B | | | V | |
|-----|--------|-----|-------|--------|-----|-------|--------|-----|
| MJD | Period | WWZ | MJD | Period | WWZ | MJD | Period | WWZ |
| 6180 | 83 | 210 | 6210 | 90 | 224 | 6170 | 91 | 478 |
| 7170 | 90 | 171 | 7200 | 76 | 86 | 7180 | 68 | 44 |
| 7840 | 75 | 327 | 7850 | 78 | 220 | 7350 | 66 | 43 |
| — | — | — | — | — | — | 7620 | 52 | 41 |
| — | — | — | — | — | — | 7820 | 83 | 171 |
| 8320 | 82 | 112 | 8400 | 75 | 180 | 8400 | 85 | 123 |
| 8590 | 163 | 65 | — | — | — | — | — | — |
| 8980 | 102 | 400 | — | — | — | — | — | — |
| 9160 | 99 | 270 | 9060 | 98 | 171 | 9160 | 98 | 80 |
| — | — | — | 9360 | 85 | 46 | 9380 | 94 | 66 |
| — | — | — | — | — | — | 10150 | 61 | 158 |
| 10490 | 83 | 254 | 10510 | 85 | 114 | — | — | — |
| 10930 | 217 | 140 | 10920 | 212 | 158 | 10920 | 208 | 139 |
| 11220 | 122 | 112 | 11250 | 123 | 93 | 11390 | 123 | 92 |
| 11400 | 123 | 110 | 11410 | 123 | 108 | — | — | — |

Table 4. Peak periods observed in the UBV WWZ transforms roughly grouped by date. Dates have been rounded to the nearest 10. Periods and WWZ output are rounded to integer values. MJD = JD – 2440000, cont.

| | U | | | B | | | V | |
| MJD | Period | WWZ | MJD | Period | WWZ | MJD | Period | WWZ |
|---|---|---|---|---|---|---|---|---|
| 11690 | 175 | 87 | 11670 | 172 | 86 | 11680 | 172 | 70 |
| 12260 | 88 | 350 | 12220 | 89 | 128 | 12240 | 90 | 81 |
| 12550 | 119 | 270 | 12560 | 121 | 172 | 12540 | 120 | 147 |
| 12960 | 76 | 98 | 12930 | 75 | 126 | 12880 | 77 | 108 |
| 13250 | 159 | 123 | 13320 | 149 | 99 | 13270 | 150 | 62 |
| 13740 | 69 | 180 | 13770 | 69 | 150 | 13750 | 69 | 71 |
| — | — | — | 14260 | 208 | 69 | 14290 | 204 | 63 |
| 14300 | 139 | 117 | 14540 | 156 | 81 | 14540 | 156 | 71 |

Table 5. Predicted dates of stable pulsation features and their periods based upon extrapolation of trends discussed in section 4.

| | Upper Track | | | Lower Track | |
| JD | Period | Observed? | JD | Period | Observed? |
|---|---|---|---|---|---|
| 2396465 | 340 | | 2397831 | 249 | |
| 2399747 | 325 | | 2401120 | 239 | |
| 2403029 | 310 | | 2404409 | 228 | |
| 2406311 | 295 | | 2407698 | 218 | |
| 2409593 | 280 | | 2410987 | 207 | |
| 2412875 | 266 | | 2414276 | 196 | |
| 2416157 | 251 | | 2417565 | 186 | |
| 2419439 | 236 | | 2420854 | 175 | |
| 2422721 | 221 | | 2424143 | 165 | |
| 2426003 | 206 | | 2427432 | 154 | |
| 2429285 | 192 | | 2430721 | 144 | |
| 2432567 | 177 | | 2434010 | 133 | |
| 2435849 | 162 | | 2437299 | 123 | |
| 2439131 | 147 | | 2440588 | 112 | |
| 2442413 | 132 | | 2443877 | 102 | |
| 2445695 | 118 | | 2447166 | 90.0 | Y |
| 2448977 | 102.7 | Y | 2450492 | 82.7 | Y |
| 2452259 | 87.9 | Y | 2453744 | 68.9 | Y |
| 2455541 | 73 | | 2457026 | 66 | |
| 2458823 | 58 | | 2460308 | 52 | |
| 2462105 | 43 | | 2463590 | 37 | |
| 2465387 | 29 | | 2466872 | 22 | |
| 2468669 | 14 | | 2470154 | 7 | |

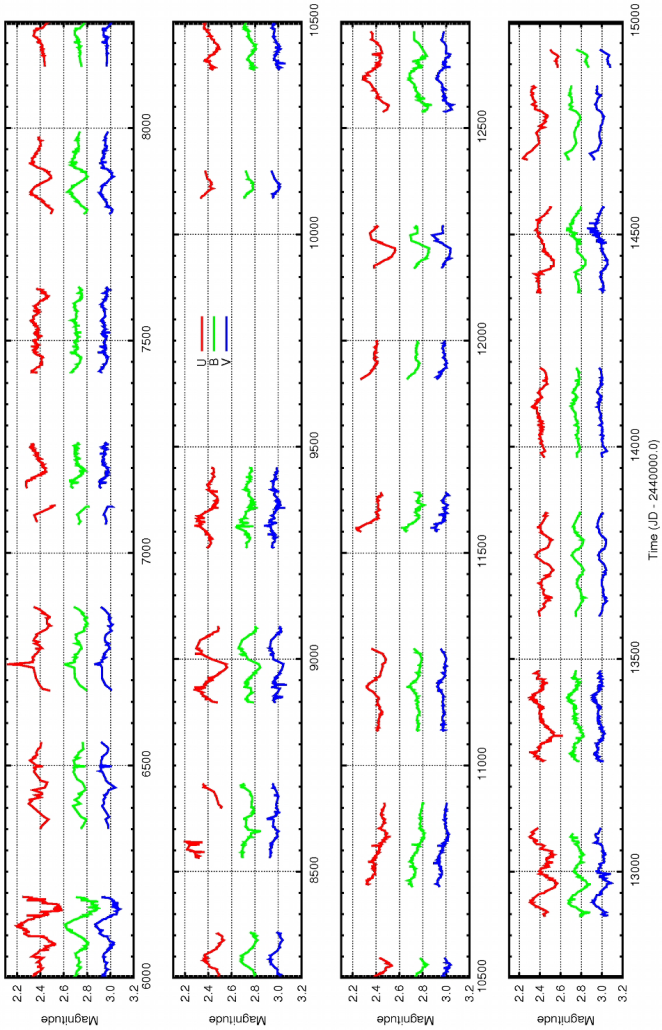

Figure 1. Inter-eclipse *UBV* photometry of ε Aur, JD 2446000–2455000 (1984 October 26–2009 June 17) from the Boyd and Hopkins data sets. For display purposes, we have offset the *U* data by –1.3 magnitude (that is, $U_{plot} = U_{data} + (-1.3)$). Likewise, the *B* data were offset by –0.8 magnitude. The V data are unaltered. Internal uncertainties (~10 mmag) are about the thickness of the lines. The observing windows consist of about 200 days of data followed by 165-day gaps. During the interval 2449500–2450000 the APT-10 photometer was not operational. The out-of-eclipse variations have amplitudes of ~0.1 magnitude in *U* and ~0.05 in *V* with characteristic timescales of 60–100 days. The sudden brightening seen around JD 2446736–2446737 might have been a flare.

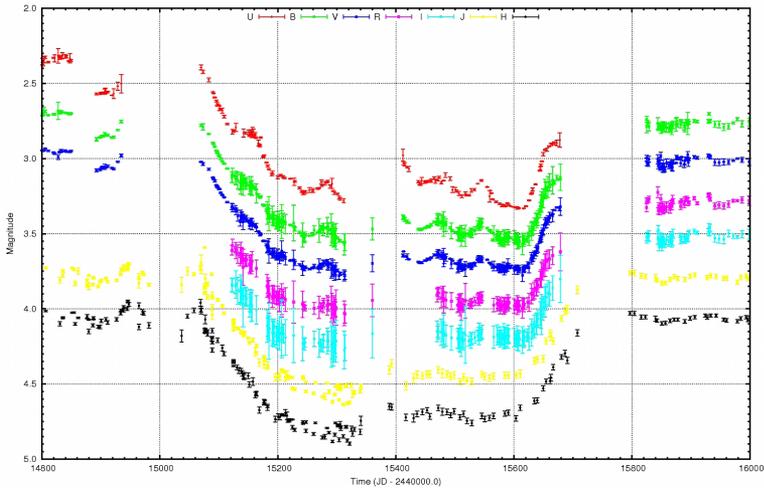

Figure 2. 2009–2011 eclipse of ε Aur in *UBVRIJH* filters, JD 2454800–2456000 (2008 November 29–2012 March 13), as measured by Hopkins (*UBV*), the AAVSO BSM (*BVRI*), and AAVSO observers Brian McCandless and Thomas Rutherford (*JH* data). The *V*-band data are plotted as observed, all other filters have been offset by an arbitrary amount for display purposes. The eclipse may be represented by a linear decrease in brightness of ~0.7 magnitude, followed by a flat minimum and then a sharp rise back to out-of-eclipse brightness. The out-of-eclipse variations are superimposed on this profile and result in 60- to 100-day cycles with characteristic amplitudes of ~0.1 magnitude in *U*, decreasing in amplitude towards longer wavelengths. Notice during the second half of the eclipse the *U*-band light curve slopes downward, whereas the *H*-band has an upward slope. This attests to the fact that the eclipse had wavelength-dependent extinction.

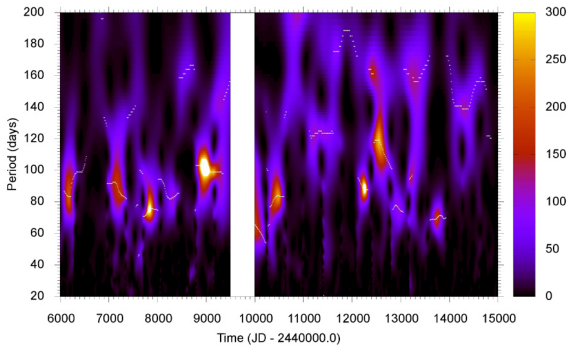

Figure 3. *U*-band WWZ period analysis of ε Aur (c = 1.25E–2). The color indicates power associated with a given period at a particular time, whereas the white dot traces the dominant period at a particular time. The region JD 2449500–2450000 has been blocked out due to a lack of data. We caution the reader that periods within < 200 days of this region may not be trustworthy. See section 4 for a discussion of the periods present and our interpretations.

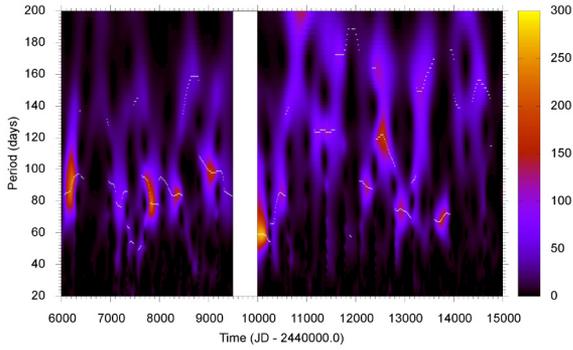

Figure 4. *B*-band WWZ period analysis of ε Aur (c = 1.25E–2). See description in Figure 3.

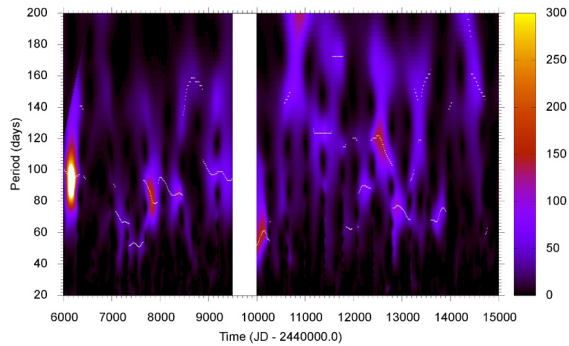

Figure 5. *V*-band WWZ period analysis of ε Aur (c = 1.25E–2). See description in Figure 3.

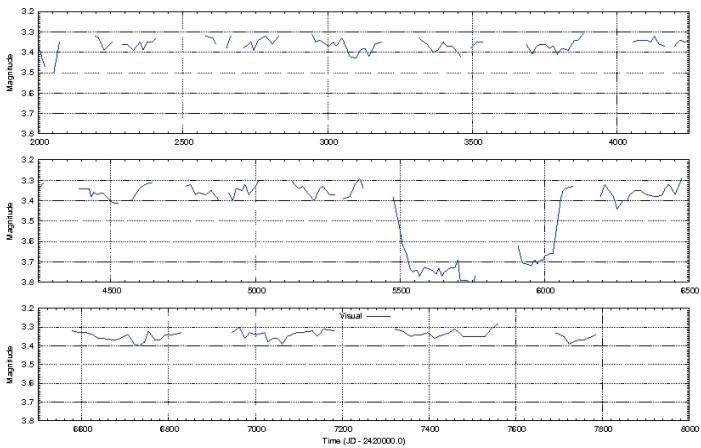

Figure 6. Visual photometry of ε Aur by Plassman (in Güssow 1936) showing the interval JD 2422000–2428000 (1919 February 10–1935 July 16) including the 1927 eclipse. Uncertainties, not plotted, are ±1 in the least significant digit (± 0.01 magnitude). The OOE variations are clearly present. Typical peak-to-peak times are difficult to judge, except right before the 1927 eclipse, where 330- to 370-day periods appear to be present.